\documentclass[a4paper]{jpconf}
\usepackage{graphicx}
\usepackage{axodraw}
\usepackage{cite}

\begin{document}
\title{Automated computation of scattering amplitudes}

\author{Giovanni Ossola}

\address{Physics Department, New York City College of Technology, The City University of New York,
300 Jay Street Brooklyn, NY 11201, USA}
\address{The Graduate School and University Center, The City University of New York,
365 Fifth Avenue, New York, NY 10016, USA}

\ead{gossola@citytech.cuny.edu}

\begin{abstract}
We review some of the recent advances in the computation of one-loop scattering amplitudes
which led to the construction of efficient and automated computational tools for NLO predictions. Particular attention is devoted to unitarity-based methods and integrand-level reduction techniques. 
Extensions of one-loop integrand-level techniques to higher orders are also briefly illustrated.
\end{abstract}

\newcommand{\txblu}{ }

\newcommand{\gnote}[1]{[\underline{Gio:} \textit{#1}  ] }

\newcommand{\beq}{\begin{equation}}
\newcommand{\eeq}{\end{equation}}
\newcommand{\bqa}{\begin{eqnarray}}
\newcommand{\eqa}{\end{eqnarray}}
\newcommand{\bite}{\begin{itemize}}
\newcommand{\eite}{\end{itemize}}
\newcommand{\bd}{\begin{displaymath}}
\newcommand{\ed}{\end{displaymath}}
\def\db#1{\bar D_{#1}}
\def\zb#1{\bar Z_{#1}}
\def\d#1{D_{#1}}
\def\tld#1{\tilde {#1}}
\def\slh#1{\rlap / {#1}}
\newcommand{\nn}{\nonumber}
\newcommand{\nl}{\nonumber \\}
\def\eqn#1{Eq.~(\ref{#1})}

\def\Formcalc{{{\sc FormCalc}}}
\def\Feynarts{{{\sc FeynArts}}}
\def\Gosam{{{\sc GoSam}}}
\def\gosam{{{\sc GoSam}}}
\def\mathematica{{{\sc mathematica}}}
\def\sam{{{\sc S@M}}}
\def\form{{{\sc form}}}
\def\blackhat{{{\sc BlackHat}}}
\def\openloops{{{\sc OpenLoops}}}
\def\madloop{{{\sc MadLoop}}}
\def\madgraph{{{\small \sc MadGraph}}}
\def\herwig{{{\sc Herwig}}}
\def\madevent{{{\small \sc MadEvent}}}
\def\helacnlo{{{\sc Helac-NLO}}}
\def\helac{{{\sc Helac}}}
\def\recola{{{\sc recola}}}
\def\ngluon{{{\sc NGluon}}}
\def\njet{{{\sc NJet}}}
\def\rocket{{{\sc Rocket }}}
\def\samurai{{{\sc samurai}}}
\def\Sherpa{{{\sc Sherpa}}}
\def\Amegic{{{\sc Amegic}}}
\def\cuttools{{{\sc CutTools}}}
\def\C++{{{\sc c++}}}
\def\Powheg{{{\sc Powheg}}}
\def\MCFM{{{\sc MCFM}}}
\def\QCDLoop{{{\sc QCDLoop}}}
\def\OneLoop{{{\sc OneLoop}}}
\def\Golem{{{\sc Golem95C}}}
\def\FastJet{{{\sc FastJet}}}
\def\Ninja{{{\sc Ninja}}}
\def\amcnlo{{a{\sc MC@NLO}}}

\section{Introduction}

The evaluation of scattering amplitudes allows us to test the theoretical models and compare their phenomenological prediction with the results of the experiments at particle colliders.
The understanding of the structure of scattering amplitudes provides the theoretical framework to develop
new techniques for their evaluation, and ultimately to design more efficient computational algorithms for the production of physical cross sections and differential distributions.  

Theory predictions play a fundamental role in the particle physics experiments at current 
hadron colliders. The high luminosity accumulated by the experimental collaborations at the Large Hadron Collider (LHC), allowed for  a very detailed investigation of the Standard Model of particle physics. In these analyses, for example to study the properties of the recently discovered Higgs boson~\cite{Aad:2012tfa,Chatrchyan:2012ufa}, theoretical predictions are not only needed for the signal, but also for the modeling of the relevant background processes, which share similar experimental signatures~\cite{Dittmaier:2011ti, Dittmaier:2012vm, Heinemeyer:2013tqa}. Further, precise theory predictions are important in order to constrain model parameters in the event that a signal of New Physics is detected. 

The scope of this review talk\footnote{Presented at the ``International Workshop on Advanced Computing and Analysis Techniques in Physics Research'' (ACAT2013), Beijing, China, May 2013.} 
is to summarize the recent progress in the evaluation of scattering amplitudes, 
which led to the development of powerful automated computational tools for Next-to-Leading Order (NLO) 
calculations. After a general overview of the many different strategies which are currently available 
for the evaluation of one-loop scattering amplitudes, in particular for the calculation of  
the virtual part, we will focus on the description of integrand-level techniques. 
The extensions of the integrand-level approach to higher orders in perturbation theory 
will also be briefly illustrated. 

The study of scattering amplitudes has a long tradition and an extensive literature, which would be difficult to condense in one presentation. 
We refer the interested reader to detailed and comprehensive reviews~\cite{Bern:2007dw, Ellis:2011cr} for a more complete picture of the field.   

\section{Scattering Amplitudes at Next-to-Leading Order}

In order to properly describe the data collected by the experimental collaborations 
at the LHC, theory predictions are not reliable without accounting for higher orders. Leading-order (LO) 
results, which for most processes can be obtained with a tree-level calculation, are affected by
large theoretical errors. 

There are two main sources of theoretical uncertainties: the \emph{truncation error}, which is related to the size of the unknown missing terms due to the truncation of the perturbative expansion, and the \emph{parametric error}, that depends on the uncertainty on the various input parameters which enter in the theoretical predictions.
The most common strategy employed to assess the theoretical error is to study the dependence of the predictions upon variation of the unphysical factorization and renormalisation scales. 

A further complication is represented by the fact that, in order to obtain results for hadronic collisions, 
the partonic matrix element (hard scattering process) should be convoluted with the parton distribution functions (PDF), which represent the low momentum scale dynamics of the partons inside the hadron. Such convolution is performed over the fractional momentum carried by the partons. While the hard scattering cross section can be obtained in perturbative QCD, the PDFs are obtained by fitting the data coming from a variety of experiments (deep-inelastic scattering, DrellÐYan, jet production) and represent a very sizable source of theoretical uncertainty in the final results for hadronic cross sections and distributions~\cite{Alekhin:2011sk}.

Tree-level cross sections suffer from a large scale dependence, which is contained in the 
running of the couplings. The scale dependence is further enhanced by the presence of multi-leg processes, i.e. additional jets in the final state, because of the insertion of additional powers of the coupling constants.
The situation is substantially improved with the inclusion of NLO effects. 

Aside from providing a reduction in the theoretical uncertainties, NLO corrections are often sizable, 
in particular in QCD. Moreover, for some processes, NLO contributions have a big effect on the shape of the the differential distributions. 

\subsection{The structure of NLO calculations}
 
The computation of NLO matrix elements requires, in addition to the tree-level LO result, the evaluation of one-loop virtual corrections (virtual part), obtained adding a virtual particle to the LO diagrams, and contributions from real emission (real part), obtained by adding one additional particle in the final states. Both terms are separately divergent: the virtual part is both ultraviolet (UV) and infrared (IR) divergent, and the UV divergence is removed by the renormalization procedure;  the real part is also IR divergent due to the presence of soft and collinear singularities. Only their combination leads to a finite physical result, since the IR divergences in the virtual part are cancelled by the ones which appear in the real part. 

The cancellation among the IR singularities in the real and virtual parts involves integrals over different phase-space configurations. This problem can be solved by employing a subtraction method~\cite{Frixione:1995ms, Catani:1996vz, Kosower:1997zr, Catani:2002hc, Nagy:2007ty, Chung:2010fx}, namely by including a subtraction term which has the same singularity structure as the real-emission corrections and cancels the infrared divergences in the phase space integral. The same contribution is then added to the virtual term, after integrating out the additional particle. After carrying out this procedure, the phase-space integrals are infrared-safe and their evaluation can be performed numerically in four dimensions. Several automated programs, based on different subtraction methods, are currently available to perform this task~\cite{Gleisberg:2007md,Seymour:2008mu,Frederix:2008hu,Czakon:2009ss,Frederix:2009yq,Hasegawa:2009tx,Frederix:2010cj, Bevilacqua:2013iha}.  

While the LO matrix elements and the NLO real parts have been available for a long time, thanks to
powerful tools for tree-level computations~\cite{Stelzer:1994ta, Kanaki:2000ey, Mangano:2002ea, Corcella:2002jc, Maltoni:2002qb, Gleisberg:2003xi, Boos:2004kh,Pukhov:2004ca, Gleisberg:2008ta, Alwall:2007st, Kilian:2007gr, Cafarella:2007pc}, until recently the evaluation of the virtual part of one-loop contributions represented the bottleneck towards the automation of NLO computation.

\section{Calculation of NLO virtual corrections}

The standard method for the computation of NLO virtual corrections relies on the evaluation of all
the Feynman diagrams associated with the process. The general task of the calculation is to compute, for each diagram contributing to the amplitude and for each phase space point, the following integral:
\beq \label{integr}
{\cal M} = \int d^n {\bar q} \,\, A(\bar q) = \int d^n {\bar q} \frac{N(\bar q)}{\db{0} \db{1} \ldots \db{m-1}}\, .
\eeq

It is well known  that the evaluation of the one-loop diagrams can be performed by decomposing each integral 
${\cal M}$ in terms of a finite set of scalar master integrals (MI)~\cite{Passarino:1978jh, 'tHooft:1978xw}.  The traditional one-loop ``master'' formula
\bqa  \label{master}
{\cal M} &=& \sum_{i_0 < i_1 < i_2 < i_3}^{m-1}
         {\bf  d( i_0 i_1 i_2 i_3 ) } \int d^n {\bar q} \, \frac{1}{\db{i_0} \db{i_1} \db{i_2} \db{i_3}}  
     + \sum_{i_0 < i_1 < i_2 }^{m-1}
         { \bf c( i_0 i_1 i_2)} \int d^n {\bar q} \, \frac{1}{\db{i_0} \db{i_1} \db{i_2} } \nl 
     &+&
\sum_{i_0 < i_1 }^{m-1}
         {\bf b(i_0 i_1)} \int d^n {\bar q}\,  \frac{1}{\db{i_0} \db{i_1}}  
     + \sum_{i_0}^{m-1}
          {\bf a(i_0)} \int d^n {\bar q} \, \frac{1}{\db{i_0} } 
     +  {\cal R}
\eqa 
allows one to express any one-loop integral in terms of 4-, 3-, 2- and 1-point scalar integrals, plus an additional rational function, known in the literature as \emph{rational part} ${\cal R}$, which is a function of the masses and momenta appearing in the original amplitude, and does not contain any logarithm or poly-logarithm.
The expressions for the finite parts and the UV and IR poles of all scalar integrals are well known, and have been codified in publicly available libraries for their numerical evaluation~\cite{vanOldenborgh:1990yc, Hahn:1998yk, Ellis:2007qk, vanHameren:2010cp, Cullen:2011kv}.

In general, each one-loop calculation consists of  three main phases:
i) the \emph{generation} of the amplitudes, namely the evaluation of the unintegrated amplitudes ${\cal M}$, in particular their numerator functions  $N(q)$ and the set of corresponding denominators;
ii) the \emph{reduction} of the amplitude to scalar MIs, whose task is to determine the 
all coefficients appearing in front of the scalar integral in the master formula  of \eqn{master} and the rational term ${\cal R}$; 
iii) the \emph{evaluation} of the MIs, which multiplied by the coefficients provide the final result. 

The construction of a similar pattern for the evaluation of the virtual contributions in higher-order calculations, starting from the determination of a general basis for multi-loop amplitudes, namely the multi-loop equivalent of \eqn{master}, have been the subject of several studies. We will briefly return to this topic in the last part of this presentation.

The analytic expressions of the coefficients in \eqn{master}, as well as the rational term ${\cal R}$, can be extracted in a fully algebraic way, separately for each process. 
This procedure, known as algebraic tensor reduction, works well for processes with a small number of particles, but was limited for multi-leg applications by its algebraic complexity and the appearance of spurious singularities.
Nevertheless, improved tensorial reduction methods~\cite{Denner:2005nn,Binoth:2008uq, Heinrich:2010ax,Fleischer:2010sq} led to the development of tools that are able to deal efficiently and precisely with processes of high complexity such as, for example,  $pp\to t\bar t b\bar b$~\cite{ Bredenstein:2009aj, Bredenstein:2010rs}.

The knowledge of the general and process-independent form of any one-loop calculation, contained in \eqn{master}, allowed for the development of new numerical and semi-numerical approaches that aim at the direct evaluation of all the coefficients without performing any algebraic reduction. 
A new powerful framework for one-loop calculation was developed by merging the idea of employing 
four-dimensional unitarity-cuts, which allow to explore the (poly)logarithmic structure of the amplitudes, contained in the cut-constructible term \cite{Bern:1994zx,Britto:2004nc}, with 
the knowledge of the universal four-dimensional decomposition for the numerator of 
the integrand for any one-loop scattering amplitudes~\cite{delAguila:2004nf} contained in the so-called OPP method~\cite{Ossola:2006us, Ossola:2007bb ,Ossola:2008xq}. 

The only part of the one-loop decomposition in Eq.~(\ref{master}) that cannot be computed using four dimensional techniques is the rational part $R$.  As is well known, even starting from a perfectly finite tensor integral, the tensor reduction may lead to integrals that need to be regularized. In dimensional regularization, this is achieved by upgrading the integration momentum to dimension $d = 4 - 2 \epsilon$, both in the numerator function and in the set of denominators. Such procedure is responsible for the appearance of the rational part ${\cal R}$.  The reconstruction of this term can be achieved by using \emph{ad hoc} tree-level Feynman rules~\cite{Ossola:2008xq,Draggiotis:2009yb,Garzelli:2009is, Garzelli:2010qm, Garzelli:2010fq, Shao:2011tg, Shao:2012ja,Page:2013xla}, by direct computation via tensorial reduction techniques~\cite{Xiao:2006vr, Binoth:2006hk},
or by the bootstrapping method~\cite{Bern:2005cq, Badger:2007si}. Other techniques employ $d$-dimensional cuts~\cite{Britto:2008sw, Britto:2008vq}, and evaluate the rational term either by performing a mass continuation generated by the additional components of the loop momentum~\cite{Bern:1995db, Anastasiou:2006jv, Anastasiou:2006gt, Badger:2008cm}, or by comparing the results obtained with cuts in two different integer dimensions~\cite{Giele:2008ve}.

Very significant improvements were achieved with the $d$-dimensional extension of unitarity methods~\cite{Ellis:2007br, Giele:2008ve,Ellis:2008ir}. The integrand-level decomposition performed in $d$ dimensions, rather than in four, exposes a richer polynomial structure of the integrand (see Section~\ref{intlev}) and allows for the combined determination of both cut-constructible and rational terms at once~\cite{Giele:2008bc, Mastrolia:2010nb}.

Aside from a deeper understanding of the structure of scattering amplitudes, 
unitarity-based methods and integrand-level reduction techniques provided the theoretical framework for  development of efficient computational algorithms for NLO calculations in perturbation theory, which have been implemented in various automated codes, such as \rocket~\cite{Giele:2008bc}, \blackhat~\cite{Berger:2008sj}, 
\Formcalc~\cite{Hahn:2000kx, Agrawal:2012cv}, \helacnlo~\cite{vanHameren:2009dr, Bevilacqua:2011xh}, \madloop~\cite{Hirschi:2011pa}, 
\Gosam~\cite{Cullen:2011ac}, \openloops~\cite{Cascioli:2011va},  \recola~\cite{Actis:2012qn}, \njet~\cite{Badger:2010nx,Badger:2012pg}. The technical features of some of these tools, together with recent advances and calculations, have been described in dedicated talks~\cite{Nejad:2013ina, Cullen:2013cka, Bern:2013pya, acatPhilipp, acatValery} during this conference.

Driven by the progress in the technical treatment of both tensorial and unitarity-based algorithms for one-loop corrections and motivated by requirements of the LHC analyses, an impressive list of calculations have been preformed~\cite{Bevilacqua:2009zn, Lazopoulos:2008ex, Giele:2009ui, vanHameren:2009vq,Berger:2009zg,Berger:2009ep,Ellis:2009zw,KeithEllis:2009bu, Berger:2010vm, Bevilacqua:2010qb,Denner:2010jp,Melia:2011dw,Bevilacqua:2011aa,Greiner:2011mp,Denner:2012yc, Greiner:2012im,Cullen:2012eh,Campanario:2012bh,Bevilacqua:2012em,Gehrmann:2013aga,Campanario:2013qba,Campanario:2013mga,Bevilacqua:2013taa, Greiner:2013gca,vanDeurzen:2013rv} in the past few years. Among the most challenging recent results at NLO QCD accuracy, let us mention the first computations of cross sections for processes with seven and eight external particles, namely the production of an electroweak gauge boson in association with four~\cite{Berger:2010zx,Ita:2011wn} and five jets~\cite{Bern:2013gka} jets, provided by the {\blackhat} collaboration, and the production, in the infinite top-mass approximation, of a Higgs boson in association with three jets in gluon-fusion~\cite{Cullen:2013saa}, which contains over ten thousand diagrams including hexagons of rank seven, and was performed with  {\Gosam}. Other important calculations were recently completed: the NLO QCD corrections to Higgs boson production in association with a top quark pair and a jet~\cite{vanDeurzen:2013xla}, which involve internal massive top quarks; the NLO QCD predictions for the production of a photon pair in association with two jets~\cite{Gehrmann:2013bga}, which is an important background for Higgs boson production; the production of four~\cite{Badger:2012pf} and five 
jets~\cite{Badger:2013yda} in hadronic collisions at NLO in massless QCD; the complete results at NLO QCD accuracy for electroweak Higgs production, including vector boson fusion and Higgs-strahlung type contributions,  in association with three jets~\cite{Campanario:2013fsa}.
  
The automated computation of physical observables at NLO accuracy, such as cross sections and differential distribution, requires to incorporate the one-loop results for the virtual amplitudes within a Monte Carlo framework (MC), such as \Sherpa~\cite{Gleisberg:2008ta}, \Powheg~\cite{Nason:2004rx, Frixione:2007vw,Alioli:2010xd}, \madgraph-\madevent~\cite{Maltoni:2002qb,Alwall:2011uj}, \herwig~\cite{Arnold:2012fq}, or \amcnlo~\cite{Pittau:2012fn},  that can take care of the phase-space integration, and of the combination of the different pieces of the calculation. 
In several recent applications~\cite{Campbell:2012am, Hamilton:2012rf, Luisoni:2013cuh, Hoeche:2013mua, Cascioli:2013gfa, Cascioli:2013era}, the MC framework also provides the possibility of merging multiple NLO parton-level matrix elements with parton showers~\cite{Hamilton:2012np, Hoeche:2012yf,Gehrmann:2012yg,Frederix:2012ps,Lonnblad:2012ix}.

%

In order to facilitate the communication between the programs computing virtual one-loop amplitudes and the MC frameworks, a standard interface has been worked out during the workshop ``Physics at TeV Colliders'' at Les Houches in June 2009, called the \emph{Binoth Les Houches Accord} (BLHA)~\cite{Binoth:2010xt}, recently updated in Ref.~\cite{Alioli:2013nda}.

Within the BLHA, the interaction between the One-loop Program (OLP) and the Monte Carlo framework (MC) 
proceeds in two phases. During the first phase, called pre-runtime phase, the two programs agree on the process and settings for the computation and make all necessary preparation work. In the pre-runtime phase, the MC creates an order file, which contains information about the setup and the subprocesses it will need
from the OLP in order to  perform the computation. The OLP reads the order file and checks availability for each item. Then it returns a contract file telling the MC what it can provide.
In the second stage, after the contract has been ``signed'',  the MC requires from the OLP  the value of the virtual one-loop amplitude at specific phase-space points.  

\section{The Integrand-level approach} \label{intlev}

The reduction at the integrand level is  based on the idea of expressing the numerator 
function of the amplitude in terms of the propagators that depend on the integration momentum, in order to identify before integration the structures that will generate the scalar integrals and their coefficients and those that will vanish upon integration of the loop momentum. 

In this approach, the coefficients in \eqn{master} are numerically determined by solving a
system of algebraic equations that are obtained by: i) the numerical evaluation of the
numerator of the integrand at explicit values of the loop-variable; ii) and the knowledge
of the most general polynomial structure of the integrand itself. 

The solution of this system of equations becomes particularly simply if we evaluate the expressions for the numerator functions at the set of complex values of the integration momentum for which a given set of inverse propagators vanish, namely the integration momenta corresponding
to the so-called quadruple, triple, double, and single cuts. This feature establish a strong connection between the integrand-level techniques and generalized unitarity methods, where the on-shell conditions are imposed at the integral level.

The strength of the method lies in the fact that the only information required
in order to extract the coefficient of the MIs is the knowledge of the numerical value of
the numerator function for a finite set of values of the integration momentum, that correspond to complex
poles of the denominators. 

\subsubsection{Integrand-level Reduction in four dimensions}  \label{oppintlev}

The integral-level reduction algorithm for one-loop scattering amplitudes, also known as OPP method, was originally developed in four dimensions~\cite{Ossola:2006us, Ossola:2007bb ,Ossola:2008xq}. 
According to this approach, the numerator function $N(q)$  which appear in the integrand for any one-loop scattering amplitudes has a universal mathematical structure, independent from the particular process at hand.
 
Any four-dimensional numerator function $N(q)$ can be written, in terms of $4$-dimensional denominators
$\d{i} = ({q} + p_i)^2-m_i^2$, as:
{\footnotesize \bqa \label{oppmaster}
N(q) &=&
\sum_{i_0 < i_1 < i_2 < i_3}^{m-1}
\left[
          {\bf d( i_0 i_1 i_2 i_3 )} +
     \tld{d}(q;i_0 i_1 i_2 i_3)
\right]
\prod_{i \ne i_0, i_1, i_2, i_3}^{m-1} \d{i} +
\sum_{i_0 < i_1 < i_2 }^{m-1}
\left[  {\bf c( i_0 i_1 i_2)} +
     \tld{c}(q;i_0 i_1 i_2)
\right]
\prod_{i \ne i_0, i_1, i_2}^{m-1} \d{i} \nl
     &+&
\sum_{i_0 < i_1 }^{m-1}
\left[
          {\bf b(i_0 i_1)} +
     \tld{b}(q;i_0 i_1)
\right]
\prod_{i \ne i_0, i_1}^{m-1} \d{i} \,\,
     +
\sum_{i_0}^{m-1}
\left[
           {\bf a(i_0) }+
     \tld{a}(q;i_0)
\right]
\prod_{i \ne i_0}^{m-1} \d{i} \, .
\eqa }
The quantities $\tld{d}$, $\tld{c}$, $\tld{b}$, $\tld{a}$, that still depend on the integration momentum $q$, are called ``spurious terms'' because they vanish upon integration and do not contribute to the final result for the scattering amplitude. Their functional form, namely their expression in terms of the integration momentum $q$, is process-independent and it is provided in Ref.~\cite{Ossola:2006us}.
Inserted back in \eqn{integr}, this expression simply states the multi-pole nature of any $m$-point one-loop amplitude. The quantities $d$, $c$, $b$, and $a$, indicated in boldface characters in \eqn{oppmaster}, do not depend on $q$ and are exactly the set of coefficients which appear in front  of the 4-, 3-, 2-, and  1-point one-loop scalar functions of \eqn{master}.

Once  \eqn{oppmaster} is established,  the task of computing the one-loop amplitude is reduced to the algebraic problem of extracting all the coefficients by evaluating the function $N(q)$ a sufficient number of times at different values of $q$. This is achieved very efficiently if we employ values of $q$ such that a subset of denominators $\d{i}$ vanish: such values correspond to the so-called quadruple, triple, double, and single cuts also used in the unitarity-cut method. 
Operating in this manner, the system of equations becomes triangular.
First one determines all the coefficients of the 4-point functions,
then moves on to the 3-point coefficients and so on.
%

The technique described above allows for the determination of the cut-constructible part, which can be fully achieved in four dimensions. 
The calculation of rational term ${\cal R}$ can be split in two separate parts, which have different origins. A first set ${\cal R}_1$ appears from the mismatch between the $d$-dimensional denominators of the master scalar integrals and the $4$-dimensional denominators of  \eqn{oppmaster}. 
The term ${\cal R}_1$ can be recovered automatically by evaluating the amplitudes for a shifted value of the mass~\cite{Ossola:2006us}. A second set ${\cal R}_2$ comes from the $d$-dimensionality of the numerator function, and can be  recovered by means of \emph{ad hoc} tree-level-like Feynman rules, that are provided in Refs.\cite{Ossola:2008xq,Draggiotis:2009yb,Garzelli:2009is,Garzelli:2010qm, Garzelli:2010fq, Shao:2011tg, Shao:2012ja,Page:2013xla} for different models.

The four-dimensional integrand-level reduction algorithm has been implemented in the code {\cuttools}~\cite{Ossola:2007ax},  that is publicly available.  The method itself does not provide specific recipe for the generation of the numerator function.
Some of the early calculations based on {\cuttools} that appeared in the literature~\cite{Binoth:2008kt,Actis:2009uq} employ traditional Feynman diagrams for the generation of the amplitudes. More recently, {\cuttools} has been incorporated within automated tools for the computation of NLO correction, such as \Formcalc, which generates the unintegrated amplitudes algebraically~\cite{Hahn:2009bf}, or 
{\sc Helac-Nlo} and {\sc MadLoop}, which construct numerically the one-loop $n$-particle amplitudes starting from tree-order amplitudes with $n+2$ particles~\cite{vanHameren:2009dr}.

\subsubsection{$D$-dimensional Integrand-level Reduction}  \label{dintlev}

In the context of four-dimensional techniques, the evaluation of  the cut-constructible term and the rational term, that escapes the four-dimensional detection, are necessarily performed separately.  As discussed above, the reconstruction of the latter usually requires information from an extra source.

The idea of performing unitarity-cuts in $d$-dimension was the basis for the development of a new algorithm, called {\samurai}~\cite{Mastrolia:2010nb}, which relies on the extension of the OPP polynomial structures
to include an explicit dependence on the extra-dimensional parameter $\mu$
needed for the automated computation of the full rational term according to the $d$-dimensional approach, the parametrization of the residue of the quintuple-cut in terms of the extra-dimension scale \cite{Melnikov:2010iu} and the numerical sampling of the multiple-cut solutions via Discrete Fourier Transform~\cite{Mastrolia:2008jb}.

The $d$-dimensional numerator $N({\bar q})$ can be expressed in terms of $d$-dimensional denominators 
$\db{i}$, as follows
\bqa
N({\bar q}) &=&
\sum_{i < \!< m}^{n-1}
          \Delta_{ i j k \ell m}({\bar q})
\prod_{h \ne i, j, k, \ell, m}^{n-1} \db{h} 
+\sum_{i < \!< \ell}^{n-1}
          \Delta_{ i j k \ell }({\bar q})
\prod_{h \ne i, j, k, \ell}^{n-1} \db{h} 
+ \nl     &+&
\sum_{i < \!< k}^{n-1}
          \Delta_{i j k}({\bar q})
\prod_{h \ne i, j, k}^{n-1} \db{h} 
+\sum_{i < j }^{n-1}
          \Delta_{i j}({\bar q}) 
\prod_{h \ne i, j}^{n-1} \db{h} 
+\sum_{i}^{n-1}
          \Delta_{i}({\bar q}) 
\prod_{h \ne i}^{n-1} \db{h} \ , \qquad
\label{dOPP}
\eqa
where $ i < \!< m $ stands for a lexicographic ordering $i < j < k < \ell < m$ and 
a bar denotes objects in $d=~4-2\epsilon$ dimensions, following the prescription
$\slh{{\bar q}} = \slh{q} + \slh{\mu}$  with ${\bar q}^2= q^2 - \mu^2$ . 

\eqn{dOPP} is the $d$-dimensional counterpart of \eqn{oppmaster}, where the functions $\Delta({\bar q}) = \Delta(q,\mu^2)$ are polynomials in the components of $q$ and in $\mu^2$. Their detailed expression is provided in Ref.~\cite{Mastrolia:2010nb}.
By substituting the decomposition of \eqn{dOPP} in \eqn{integr}, the multi-pole nature of the integrand of any one-loop $n$-point amplitude becomes transparent:
{\small \bqa
A(\bar q) &=&
\sum_{i < \!< m}^{n-1}
         { \Delta_{ i j k \ell m}({\bar q}) \over 
           \db{i} \db{j} \db{k} \db{\ell} \db{m} } 
+
\sum_{i < \!< \ell}^{n-1}
         { \Delta_{ i j k \ell }({\bar q}) \over 
           \db{i} \db{j} \db{k} \db{\ell} } 
+
\sum_{i < \!<  k }^{n-1}
         { \Delta_{i j k}({\bar q}) \over 
           \db{i} \db{j} \db{k} }
+ 
\sum_{i < j }^{n-1}
         { \Delta_{i j}({\bar q}) \over 
           \db{i} \db{j} } 
+
\sum_{i}^{n-1}
         { \Delta_{i}({\bar q}) \over
           \db{i} } \ .
\eqa
}

Once the $d$-dimensional identity of  \eqn{dOPP} has been established, the calculation of a generic scattering amplitude amounts to the problem of extracting the coefficients of multivariate polynomials, generated at every step of the multiple-cut analysis.

After all coefficients contained in $\Delta_{ i j k \ell m}$, $\Delta_{ i j k \ell}$, $\Delta_{ i j k}$, $\Delta_{ i j}$,  
and $\Delta_{ i }$ have been extracted, they are multiplied by the corresponding MIs.
In addition to the standard scalar integrals already contained in the 4-dimensional decomposition, 
there are additional $\mu^2$-dependent master integrals:
\bqa
\int d^n \bar{q}
\frac{\mu^2}{\db{i}\db{j}} \,\,\,\,\, , \,\,\,\,\,
\int d^n \bar{q}
\frac{\mu^2}{\db{i}\db{j}\db{k}} \,\,\,\,\, , \,\,\,\,\,
\int d^n \bar{q}
\frac{\mu^4}{\db{i}\db{j}\db{k} \db{l}} \nn \, ,
\eqa
whose expressions are also well-known~\cite{Pittau:1996ez,Bern:1995db}.
The presence of these new contributions, together with the $d$-dimensional decomposition, account for the complete evaluation of the full rational term.

The {\samurai} code~\cite{Mastrolia:2010nb} is publicly available. The method itself does not provide specific recipe for the generation of the numerator function: {\samurai} can reduce integrands defined either as {\it numerator functions} sitting on products of denominators, to be used with calculations based on Feynman diagrams, or as {\it products of tree-level amplitudes} sewn along cut-lines, to be employed for the reduction of amplitudes generated with unitarity-based techniques. 
The reduction provided by {\samurai} has been has been employed within the {\Gosam} framework, as well as interfaced~\cite{Agrawal:2011tm}  with \Formcalc.

The integrand decomposition was originally developed for renormalizable gauge theories,
where, at one-loop, the rank of the numerator cannot be greater than the number of external legs.
To deal with the presence of effective-gluon vertices generated by the large top-mass limit, required by the evaluation of $pp \to H +2,3$ jets in gluon fusion~\cite{vanDeurzen:2013rv, Cullen:2013saa}, the reduction code within {\samurai} was upgraded to accommodate an extension of the polynomial residues and of the corresponding sampling required to fit all coefficients~\cite{Mastrolia:2012du, Mastrolia:2012bu}.

\subsubsection{Integrand Reduction via Laurent Expansion}

Elaborating on the the techniques proposed in~\cite{Forde:2007mi,Badger:2008cm}, a different approach to the integrand-reduction method for one-loop amplitudes was recently presented~\cite{Mastrolia:2012bu},  which allows to extract the coefficients of the integrand decomposition by
performing a Laurent expansion, whenever the analytic form of the numerator function is
known.

In general, when the multiple-cut conditions do not fully constrain the loop momentum, the on-shell 
solutions are still functions of some free parameters, possibly the components of the momentum which are not frozen by the cut conditions.
The integrand-reduction algorithm as implemented in the numerical codes of Refs.~\cite{Ossola:2007ax,Mastrolia:2010nb} requires, in order to extract the value of the unknown coefficients, to solve a system of equations obtained by sampling the numerator on a finite set of values of such free parameters after subtracting all the non-vanishing contributions coming from higher-point residues.

The reduction algorithm can be simplified by exploiting the knowledge of the
analytic expression of the integrand.  Indeed, by performing a
\emph{Laurent expansion} with respect to one of the free parameters
which appear in the solutions of the cut,
both the integrand and the subtraction terms exhibit the same polynomial behavior of the residue.  
Moreover, the contributions coming
from the subtracted terms can be implemented as \emph{corrections at
  the coefficient level}, hence replacing the subtractions at the
integrand level of the original algorithm. 
The parametric form of this corrections can be computed once and for all, in terms of a
subset of the higher-point coefficients.  With this method the number
of coefficients entering in each subtracted term is significantly
reduced.  For instance, box and pentagons do not affect at all the
computation of lower-points coefficients.

If either the analytic expression of the integrand or the tensor structure of
the numerator is known, this procedure can also be implemented in a
semi-numerical algorithm.  Indeed, the coefficients of the Laurent
expansion of a rational function can be computed, either analytically
or numerically, by performing a \empty{polynomial division} between the
numerator and the denominator.  This method has been implemented, within the {\Gosam} framework, in the \C++ library {\Ninja}, showing an improvement in 
the computational performance, both in terms of speed and precision, with respect to the standard algorithms.
The new library has been recently employed in the evaluation of NLO QCD corrections to $p p \to t {\bar t} H j $~\cite{vanDeurzen:2013xla}.

\section{Beyond NLO}

The Next-to-Next-to-Leading-Order (NNLO) computations are quite far from automation 
and only a few computations are available for processes at hadron colliders~\cite{Czakon:2013goa, Boughezal:2013uia}.

At one-loop, the advantage of knowing that one complete basis of MIs is formed by scalar 
one-loop functions \cite{Passarino:1978jh} and the availability of their analytic expression
allowed the community to focus on the development of efficient algorithms  
for the extraction of the coefficients multiplying each MI. 
At higher-loop, a general basis of MIs is not known and they are only identified at the end of the reduction procedure. Moreover, many MIs do not have a known analytic expression and they should be evaluated numerically. 
The multi-loop reduction technique which is most often employed is the well-known Laporta algorithm~\cite{Laporta:2001dd}, based on the solution of algebraic systems of equations obtained through integration-by-parts identities~\cite{Tkachov:1981wb}. 

\subsection{Integrand-Level Techniques Beyond One-Loop}

Extensions of the integrand reduction method beyond one-loop, first proposed in Refs.~\cite{Mastrolia:2011pr,Badger:2012dp}, 
have recently become the topic of several studies~\cite{Zhang:2012ce, Mastrolia:2012an,Kleiss:2012yv, Badger:2012dv,Feng:2012bm, Mastrolia:2012wf, Mastrolia:2013kca}, thus providing a new direction in the study of multi-loop amplitudes (see also the presentation of S.~Badger at this conference~\cite{Badger:2013sta}). An alternative approach based on maximal unitarity has been developed in Refs.~\cite{Gluza:2010ws, Kosower:2011ty, Larsen:2012sx, CaronHuot:2012ab,Johansson:2013sda}.

Higher-loop extension of the integrand-level techniques require a proper parametrization
of the residues at the multi-particle poles~\cite{Mastrolia:2011pr}. 
The parametric form of the polynomial residues is {\it process-independent}
and can be determined in a general way from the corresponding multiple cut: only the values of the coefficients which appear in the residues is {\it process-dependent}.
Each residue can be written as a multivariate polynomial in the {\it irreducible scalar products} (ISP), namely 
the set of scalar products, involving one  or more the loop momenta, which cannot be reconstructed in terms of denominators. The ISPs either yield {\it  spurious contributions}, which vanish upon integration, or generate the integrals which form the basis of amplitude decomposition~\cite{Mastrolia:2011pr,Badger:2012dp}.

In Refs.~\cite{Zhang:2012ce, Mastrolia:2012an}, the general determination of the residues at the multiple cuts has been systematized within the mathematical framework of algebraic geometry as a problem of
multivariate polynomial division. The use of these techniques proved that the integrand decomposition, originally formulated for one-loop amplitudes, is applicable
at any order in perturbation theory, irrespective of the complexity of the topology of the diagrams involved, massless or massive, planar or non planar. 
The shape of the residues is uniquely determined  by the on-shell conditions, without any additional constraint. 
An iterative integrand-recursion formula, based on successive divisions of the numerators modulo the 
Gr\"obner basis of the ideals generated by the cut denominators, can provide the required multi-particle pole decomposition for arbitrary amplitudes, independently of the number of loops.
The shape of the residues is uniquely determined by the on-shell conditions, without any additional constraint. 

The algorithm presented in Ref.~\cite{Mastrolia:2012an} relies on general properties of the loop integrands, which for simplicity we write as:
 \beq
 \mathcal{I}_{i_1\cdots i_n}  = \frac{{\cal N}_{i_1\cdots i_n}}{\db{i_1} \cdots \db{i_n}} \; . 
 \label{igen}
 \eeq
i) When the number $n$ of denominators $\db{i}$ is larger than the total number of the
components of the loop momenta, namely when the on-shell conditions $\db{1}=\db{2}=\ldots=\db{n}$ have no solutions, the integrand $\mathcal{I}_{i_1\cdots i_n}$ is {\it reducible}: it  can be written in terms of integrands with $(n-1)$ denominators (lower point functions). This is the case, for example, of six-point functions in the one-loop decomposition.\\
ii) When  $n$ is equal or less than the total number of components of the loop momenta, the on-shell conditions $\db{1}=\db{2}=\ldots=\db{n}$ have solutions. To determine the corresponding residue, we divide the numerator  ${\cal N}_{i_1\cdots i_n}$ modulo the Gr\"obner basis of the $n$-ple cut, namely modulo a set of polynomials that vanish on the same on-shell solutions as the cut denominators. The {\it remainder}
of the division is the {\it residue}  $\Delta_{i_1\cdots i_n}$ of the $n$-ple cut.  The {\it quotients} 
generate integrands with $(n-1)$ denominators which should undergo the same decomposition.
This allows us to cast the each numerator ${\cal N}_{i_1\cdots i_n}$, sitting on a set of denominators $\db{i}$, in the form
\beq
{\cal N}_{i_1\cdots i_n} =
\sum_{\kappa=1}^{n}   
{\cal N}_{i_1\cdots i_{\kappa -1}i_{\kappa+1}\cdots i_n}\, \db{i_\kappa} + \Delta_{i_1\cdots i_n} \ ,
\label{recursive}
\eeq
which inserted in \eqn{igen}, provides the recurrence relation
\beq
\mathcal{I}_{i_1\cdots i_n}  =  
 \sum_{\kappa=1}^{n}   \mathcal{I}_{i_1\cdots i_{\kappa -1} i_{\kappa+1} i_n}
+ \frac{\Delta_{i_1\cdots i_n}}{\db{i_1} \cdots  \db{i_n}}  \, .
\label{decgen}
\eeq
iii) A specific set of on-shell cut conditions, labeled {\it maximum-cuts}, are defined, for each number of loops, by the maximum number of on-shell conditions which can be simultaneously satisfied by the loop momenta. The {\it Maximum Cut Theorem}~\cite{Mastrolia:2012an} ensures that the residue at the maximum-cuts is parametrized by $n_s$ coefficients, where $n_s$ is the number of solutions of the multiple cut-conditions, and therefore can {\it always be reconstructed by evaluating the numerator at the solutions of the cut}. This theorem extends at all orders the features of the one-loop quadruple-cut \cite{Britto:2004nc,Ossola:2006us}, where the only two complex solutions of the cut determine the two coefficients 
needed to parametrize the residue. 

The integrand recurrence relation of \eqn{decgen} may be applied in two ways. When the parametric form of all residues and the solutions of all possible multiple cuts are known, all the coefficients which appear in the residues can be determined by evaluating the numerator at the solutions of the multiple cuts, as many times as the number of the unknown coefficients. 
This approach has been employed at one loop in the original integrand reduction~\cite{Ossola:2006us}, and the language of multivariate polynomial division provides its natural generalization at all loops.

As a different strategy
~\cite{Mastrolia:2013kca}, the decomposition of the amplitude can be obtained analytically by successive polynomial divisions, which at each step generate the actual residues. In this approach, the reduction algorithm described above is applied directly to the actual numerator functions, without requiring the knowledge of the parametric form of all residues or the solutions of the multiple cuts.  This new technique can naturally be applied to integrands with denominators appearing with arbitrary powers, thus solving a long-standing problem within unitarity-based methods, and represents a viable starting point towards the construction of automated computational tools at higher loops.

\section{Conclusions}

There are several different approaches available for one-loop calculations. 
The \emph{generation} of the amplitudes can be performed starting with traditional Feynman diagrams, by means of recursive relations, or by gluing tree-level sub-amplitudes, as in unitarity-based methods. Aside from traditional tensorial approaches, the coefficient can be also extracted by performing an integrand-level \emph{reduction} in 4-dimensions, as in the original OPP method, or by employing a $d$-dimensional basis, which accounts automatically for the rational terms.
Finally, the integral basis of  \eqn{master}  can be upgraded to include tensorial terms, to improve stability or performance. 

The combined developments of all these different techniques, together with an increase in the performance and availability of computational facilities, triggered the ``NLO revolution''~\cite{PerretGallix:2013bg, Salam:2011bj}. The full automation of NLO calculations has been successfully achieved by means of different tools that, just like their tree-level predecessors, allow the user to compute full NLO virtual corrections at the simple effort of 
providing the list of particles
and some input parameters.  It is indeed fascinating to witness the number and the quality of advanced automated NLO calculations that have been performed in the past few years~\cite{Bern:2008ef, Binoth:2010ra, AlcarazMaestre:2012vp}.

Automated codes for the evaluation of NLO QCD and EW corrections have been also successfully interfaced within MC tools to produce results that will be of great importance  for the experimental analyses, starting with the studies of the properties of the Higgs boson and the searches for New Physics at the LHC.

\ack 
I would like to thank Gionata Luisoni and Pierpaolo Mastrolia for their valuable comments and suggestions.
This work is supported in part by the National Science Foundation under Grant No.~PHY-1068550 and PSC-CUNY Award No. 65188-00 43. 


\bibliographystyle{iopart-num}
\section*{References} 
\bibliography{references}

\end{document}